\documentclass[aps,prb,twocolumn,groupedaddress,showpacs,showkeys,citeautoscript,flushbottom,amsfonts,amsmath,amssymb]{revtex4-1}
\usepackage{graphicx}
\usepackage{nicefrac}
\usepackage[latin1]{inputenc}
\begin{document}
\small
\newcommand{\bskipdm}{\mskip -3.8\thickmuskip}
\newcommand{\bskiptm}{\mskip -3.0\thickmuskip}
\newcommand{\bskipsm}{\mskip -3.1\thickmuskip}
\newcommand{\bskipssm}{\mskip -3.1\thickmuskip}
\newcommand{\fskipdm}{\mskip 3.8\thickmuskip}
\newcommand{\fskiptm}{\mskip 3.0\thickmuskip}
\newcommand{\fskipsm}{\mskip 3.1\thickmuskip}
\newcommand{\fskipssm}{\mskip 3.1\thickmuskip}
\newcommand{\pint}{\mathop{\mathchoice{-\bskipdm\int}{-\bskiptm\int}{-\bskipsm\int}{-\bskipssm\int}}}
\newcommand{\ds}{\displaystyle}
\newcommand{\D}{\mathrm{d}}
\newcommand{\I}{\mathrm{i}}
\newcommand{\EXP}[1]{\mathrm{e}^{#1}}    

\newcommand{\tpp}{\hat{t}}
\newcommand{\rpp}{\hat{r}}
\newcommand{\zpp}{\hat{z}}
\newcommand{\ppp}{\hat{\varphi}}
\newcommand{\tbb}{\bar{t}}
\newcommand{\rbb}{\bar{r}}
\newcommand{\zbb}{\bar{z}}
\newcommand{\pbb}{\bar{\varphi}}
\newcommand{\lbb}{\bar{\ell}}
\newcommand{\ttt}{\tilde{t}}
\newcommand{\rtt}{\tilde{r}}
\newcommand{\ztt}{\tilde{z}}
\newcommand{\ptt}{\tilde{\varphi}}
\newcommand{\abs}[1]{\left\lvert #1 \right\rvert}
\newcommand{\tvv}{\check{t}}
\newcommand{\xvv}{\check{x}}
\newcommand{\omvv}{\check{\omega}}
\newcommand{\kvv}{\check{k}}
\newcommand{\vvv}{\check{v}}
\newcommand{\lvv}{\check{\ell}}
\newcommand{\pvv}{\check{\varphi}}
\newcommand{\Rom}{R_\omega}

\title{Spatial geometry of the rotating disk and its non-rotating
  counterpart}

\author{Klaus Kassner${}^{1}$} 
\affiliation{${}^{1}$Institut für Theoretische Physik, \\
  Otto-von-Guericke-Universität
  Magdeburg, Germany
}

\begin{abstract}
  A general relativistic description of a disk rotating at constant
  angular velocity is given. It is argued that conceptually this
  direct approach poses fewer problems than the special relativistic
  one. For observers on the disk, the geometry of their proper space
  is hyperbolic. This has interesting consequences concerning their
  interpretation of the geometry of a non-rotating disk having the
  same radius. The influence of clock synchronization
  on spatial measurements is discussed.
\end{abstract}
%\date{12 September 2011}
%\date{11 December 2011}
\date{23 February 2012}

\maketitle

%\sloppy
\allowdisplaybreaks
\section{Introduction}

In his famous 1916 paper introducing general
rela\-tivity,\cite{einstein16} Einstein invoked a rotating frame of
reference to give an example of how non-Euclidean geometry may arise
in relativistic physics. He considered an inertial frame and a circle,
the center of which was at rest in the frame, plus a second frame
rotating about the symmetry axis perpendicular to the circle. For
symmetry reasons, the geometric figure is a circle in both frames.
Measuring its circumference and diameter in the usual way, i.e., by
laying out rods along both lines, the ratio of the number of
rods needed in the inertial frame would approach $\pi$ using
sufficiently short rods. The same kind of measurement would produce a
number exceeding $\pi$ in the rotating frame, because due to Lorentz
contraction more rods would be needed along the circumference but not
along the diameter. Contraction arises only parallel to the direction
of motion, so radially arranged rods would not suffer from it. The
conclusion would then be that the geometry of the circle is
non-Euclidean and hyperbolic in the rotating frame.

In later discussions, the circle was replaced by a
solid disk (Ref.~\onlinecite{berenda42} is an example).
This should be quite in the spirit of Einstein, whose basic premise
was that spatial geometry is determined by the physical properties of
bodies serving as measuring devices, bodies that are approximately
rigid, within the limitations set by his theory. The circle then
becomes the circumference of the disk. In order to have a physical
realization of an ``inertial'' circle as well, one may imagine a
second non-rotating disk under the first, with exactly the same
radius. If the two disks are sufficiently close to each other, the
outline of their circumferences will essentially be a single spatial
curve, and the question is then, how observers in different states of
motion will assess the length of this curve.
 
When publishing these ideas, Einstein had already thought hard for a
number of years about the generalization of special relativity, so one
might be inclined to believe that he got his introductory example
right. And indeed, many researchers later confirmed and extended his
results.\cite{berenda42,gron75,weber97,rizzi98,rizzi99,rizzi02,ruggiero03}

However, surprisingly there are to this day controversies about the
geometry of and the physics on, a rotating
disk.\cite{pellegrini95,weber97b,klauber98,weber98,%
  tartaglia99,nikolic00,west08,semon09} The subject is still
alive, as is attested by the existence of a whole book \cite{rizzi04}
devoted to contributions by both opponents and defenders of Einstein's
results.  Some authors even consider special relativity to be
inapplicable to rotating frames \cite{selleri96,klauber98}. Since not
all of the divergences can possibly be reconciled, the reader should
be warned that there must be errors in several of these 
contributions. 
Statements range from the geometry of the rotating disk being
hyperbolic, corresponding to a relationship $L>2\pi R$ between its
circumference $L$ and its radius $R$, to the disk remaining
Euclidean\cite{klauber98,tartaglia99} ($L=2\pi R$) and the idea that
an observer traveling around a circle at constant speed should assess
its geometry to be elliptic \cite{semon09} ($L<2\pi R$).

The dispute is all the more surprising, as the assumption would seem
reasonable that a correct application of general relativity should
resolve the issue beyond any doubt.  To be sure, because gravity is
not involved, a complete description of the physics on a rotating disk
can be given within special relativity. Yet this appears to be one of
the rare cases where a direct application of general relativity
renders the problem \emph{simpler} than arguing within the special
theory. %This may have to do with the fact that the
Applicability of the general theory cannot be denied, whereas the --
mistaken -- idea still seems to be lurking in some minds that special
relativity does not work in accelerated frames.

Furthermore, general relativity, being firmly geometry-based, offers
certain conceptual advantages, improving clarity. Consider for example
the proposition by Semon et al.\cite{semon09} that an observer moving
around a circle, measuring each length element of its circumference
%passing his position 
to be Lorentz contracted, will find a total length smaller than $2\pi
R$, implying an elliptic geometry. Such an idea seems to make sense
within the framework of the special theory, where accelerated motion
has an absolute character.  The general theory, however, tells us
that acceleration is indistinguishable from the effect of an appropriate
gravitational field\cite{munoz10}, so any observer may consider
herself at rest.  Beyond that, if one cherishes the notion of a
spatial geometry that may depend on the state of motion of an
observer, which is what consideration of the rotating and static disks
suggests, then it seems obvious that in associating a space with an
observer, that space \emph{must} not move with respect to the observer
(rather than just \emph{might} not move). Otherwise we would have to
consider an infinite (three-parameter) family of spaces associated with
each observer, which would severely exacerbate the uniqueness problems
that we will see arise anyway.

Hence, what we mean by a space associated with a particular observer
is one, in which the latter is at rest.  But what could such an
observer learn about the space at large by locally measuring length
elements of a curve moving past herself?  While the moving curve
certainly contains information about distant pieces of space from the
past, we must admit that this does not help a lot, since we know from
general relativity that geometries may be time dependent, so the
information we get by measuring the curve element at the current
moment may already be outdated. Special relativity even suggests that
this is precisely what happens. At any instant in time, there is a
comoving inertial observer having the same velocity as our circle
traveler. For this inertial observer the circle is Lorentz contracted
into an ellipse along the direction of motion\cite{gron75} (so the
ellipse changes its orientation each time the circle traveler changes
her direction of motion).  The circumference of the ellipse is
different from the circumference measured by the traveler, because
length elements of the ellipse have different length contractions with
respect to the original circle, depending on their orientation. The
length element passing the traveler always has the same contraction
factor corresponding to her constant speed.  Therefore, length
elements of the ellipse change in time as they approach the traveler's
position.

Evidently, a single observer making local measurements can draw direct
conclusions only about the local piece of space that he occupies.
Observers are point-like entities in relativity. Space is an entity
defined by a collection of observers.  As we shall see later, this
statement can be given a mathematically rigorous
interpretation.\cite{rizzi02} For the time being, a simple argument
may help to convince the reader that a single observer is insufficient
to decide about the nature of a finite patch of space. The observer at
the center of our rotating disk is part both of an ensemble of
inertial observers (sitting on the non-rotating disk) and an ensemble
of accelerated observers (sitting on the rotating disk). If the
geometry is Euclidean for the inertial ensemble and non-Euclidean for
the other, this observer belongs to two different
geometries.\footnote{Similarly, \emph{any} observer at a fixed
  position on the rotating disk may be considered to belong to a
  set of instantaneously comoving inertial observers, living in
  Euclidean space, as well as to the set of stationary disk observers,
  whose space is non-Euclidean. Measurements take time, so
  the instantaneous inertial frame is not particularly
  practical for physical descriptions -- it  changes
  continuously.}  Hence, the geometry is not fixed by specifying a
single observer.

If we have to consider sets of observers to make statements about
spatial geometry, then it is useful to start the consideration from
extended solid bodies, assuming the material points of these solids to
define the rest states for local observers. Therefore, it is probably
conceptually somewhat simpler to imagine two material disks, whose
state of motion determines the characteristics of the space of
observers at rest on them, than to start from Einstein's original idea
of a circle and two frames of reference not defined in any detail.
This approach is also less problematic than the single-observer
scenario discussed in Ref.~\onlinecite{semon09}, which, as we shall
see, pertains to the question of the geometry of the non-rotating disk
in the space of observers on the rotating one, a question that rarely
has been touched upon in the literature\cite{wucknitz04}.

The remainder of this paper is organized as follows.  In
Sec.~2, two scenarios relating to the behavior of lengths of
accelerated bodies are discussed, going by the name of Bell's
spaceship paradox and Ehrenfest's paradox.  Ehrenfest's paradox is
often invoked in discussions of the rotating disk, with different
intentions.  Sometimes it is used to support the idea of a physical
shrinkage of the circumference of the disk, sometimes to demonstrate
the impossibility of uniformly rotating disks or even a failure of
special relativity in rotating frames. Bell's spaceship paradox will
help us resolve Ehrenfest's paradox and clarify the role of stresses
in possible rotating disk experiments. Experimental realizability
of the rotating disk is briefly discussed in Sec.~3.  The
formalism is developed in Sec.~4 and the proper geometry of the
rotating disk is introduced.  A description of the non-rotating disk as
observed by inhabitants of the spinning one is developed and its
synchronization dependence investigated. Some of the resulting
conclusions seem to be novel.  Finally, Sec.~5 summarizes our results
and visualizes them by way of an example.
 
\vfill

\section{Two paradoxes}
\subsection{Bell's spaceship paradox}
The thread-between-spaceships paradox\cite{dewan59} was not  actually
invented by Bell.  He just immensely contributed to its
popularization.\cite{bell76}

Imagine two spaceships moving with exactly the same acceleration
program in a given inertial system $S$ along a straight line, which we
may choose as the $x$ axis. Obviously their distance within $S$ must
remain constant by definition. Suppose a thread connects the two
spaceships. The length of the thread at the beginning of the flight is
exactly equal to the distance of the two spacecraft. Beyond that, the
thread is not assumed strong enough to withstand the thrust of the
rocket engines, should there be a conflict about the distance at which
the thread tries to keep the spaceships and the distance their
acceleration dictates. The question then is this: will the thread break
or will it not?

\begin{figure}[h!]
\includegraphics[height=4.0cm]{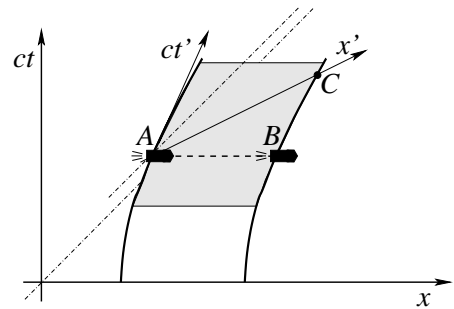}
\caption{Illustration of Bell's spaceship paradox. Dash-dotted lines
  denote world lines of light.}
\label{fig:bells_paradox}
\end{figure}

The situation is depicted in Fig.~\ref{fig:bells_paradox}, displaying
the world lines of the two spaceships in a Minkowski diagram, with the
coordinates in $S$ given as $x$ and $ct$. The shaded area is a section
of the world sheet of the thread.

At first sight, one might wonder why the question of a breaking thread
should arise at all. In any event, the distance of the spaceships
remains constant, so why should there be a problem? Well, the thread is
set in motion, so it ought to Lorentz contract. If it cannot, because
the spaceships resist its pull, it must eventually break. But a number
of relativists do not regard Lorentz contraction as ``real''. It is
considered an effect resulting from different points of view of
different observers, not a change of the object.

Bell\cite{bell76} insisted that it is real enough to make the thread
break and that in teaching relativity, this point should be driven
home with the students. In a comoving frame, the thread should keep its
proper length, if it is unstressed. This means it must be Lorentz
contracted in $S$. Therefore, if its length in $S$ is kept fixed
artificially, it will experience stresses\cite{gron81} and eventually
has to break.

Can this be rationalized from the point of view of an observer
comoving with an end point of the thread? Consider the trailing
spaceship in the picture. Its instantaneous inertial system $S'$ is
described by the axes $x'$ and $ct'$.  To obtain the axis $x'$
describing simultaneous events from the point of view of the trailing
craft, we know that we simply have to draw a straight line that makes
the same angle with the world line of a right-moving light beam as the
$ct'$ axis -- this guarantees that the world line of light is a
bisector of the time and space axes and the second postulate of
special relativity is satisfied.

If the spaceship inhabitant wishes to determine the length of the
thread, he has to measure the extension of its world sheet along a
line of simultaneity, i.e., along the $x'$ axis. This length is given
by the spatial distance of the events $A$ and $C$ of the diagram.
Because the length units along the $x$ and $x'$ axes are not normally
equal in Minkowski diagrams, we cannot easily read a quantitative
relationship off the figure. But we see that $C$ is \emph{later} on
the world line of the leading spaceship than $B$, where it had the
same speed as the trailing craft in $A$. So the leading spaceship has
a \emph{larger} velocity in $C$, an event which in the $x'\,ct'$ frame
is simultaneous with $A$.  This means that the observer at $A$ finds
the leading spaceship to increase its distance from his (and to have
done so from the beginning of the trip).  Therefore, the thread has
become longer and will break as soon as its yield limit is reached.

\subsection{Ehrenfest's paradox}
Ehrenfest's paradox\cite{ehrenfest09} has to do with the question
whether a disk can be set into rotation via Born rigid
motion.\cite{born09} From the very beginnings of special relativity,
it was clear that the theory does not permit the existence of rigid
\emph{bodies} in the sense of classical mechanics, %\cite{einstein07}
because accelerating such a body would mean transmission of a
signal instantaneously from one of its ends to the other, which is
incompatible with relativistic kinematics.

However, rigid \emph{motion} in the Born sense seemed feasible. A body
moves rigidly, if in a given inertial system each of its volume
elements is contracted in the direction of motion by the Lorentz
factor corresponding to its instantaneous velocity. Hence, its
dimensions in its own successive rest-frames are preserved, implying,
as Rindler emphasizes,\cite{rindler01} that this definition is
frame-independent. When a body is set in motion this
way, it will not experience any internal stresses (if it had none
before being moved). Translational Born rigid acceleration is indeed
possible.\cite{rindler01}

Ehrenfest considered a rotating cylinder.\cite{ehrenfest09} Assuming
the rotation to be rigid, he found two contradictory conclusions.
Radial line elements of the cylinder would not be Lorentz contracted,
being perpendicular to the local motion, so the radius $R$ of the
cylinder would be unchanged.  Hence, its circumference would be
$L=2\pi R$. On the other hand, the circumference of the cylinder would
have to be Lorentz contracted, because line elements along the rim are
aligned with the direction of local motion. Hence, the circumference
would be $L<2\pi R$. We have two incompatible results.

A solution of Ehrenfest's paradox follows from the fact that
there is no Born rigid accelerated \emph{rotation}.

\begin{figure}[h!]
\includegraphics[height=3.6cm]{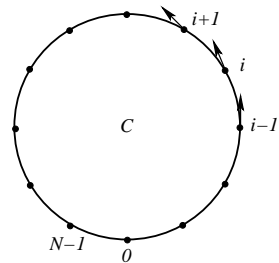}
\caption{How to make a cylinder rotate faster?}
\label{fig:ehrenfest}
\end{figure}

In order to see this, let us make use of what we have learned from
Bell's spaceship paradox. Suppose we wish to increase the rotation
speed of our cylinder by applying accelerating forces to elements of
its rim.  Figure \ref{fig:ehrenfest} gives a visualization. The cross
section of the cylinder surface is divided into $N$ equal segments. To
accelerate segment $i$, forces have to be applied at its endpoints
numbered $i-1$ and $i$. These accelerations all have to be equal due
to the rotational symmetry of the problem.  But then each segment is
in a similar situation as the thread in Bell's spaceship paradox. Its
ends are subject to the same acceleration program (as viewed by an
observer $C$ at the central axis of the cylinder), so the segment will
be longitudinally stressed, contrary to the assumption of Born rigid
motion.

Hence, material rotating cylinders will have to experience internal
stresses, if they were unstressed before rotation. This may set
practical limits on achievable rotation speeds but does not render
rotation itself impossible.

The given solution to the paradox is dynamical, invoking forces.
Gr{\o}n has shown how to solve it in a purely kinematic
way.\cite{gron75} Consider segment $i$ on the rim of the cylinder and
a comoving observer. In order to accelerate the segment further in a
way compatible with Born rigid motion, points $i-1$ and $i$ must be
accelerated (with some as yet undetermined accelerations) at the
\emph{the same time according to the comoving observer} in order to
keep the comoving length of the segment unchanged. But this means
that these accelerations have to be applied at \emph{different} times
for the central observer. If we consider the comoving direction to be
the $x'$ axis and assume a rotation speed $v$, the local Lorentz
transformations to a  frame at rest with respect to $C$ read
\begin{equation}
x = \gamma \left(x'+v t'\right)\>,  \quad
t = \gamma \left(t'+\frac{v}{c^2} x'\right)\>,
\label{eq:lorentz_back}
\end{equation}
with 
$
\gamma=1\Big/ \!\!\sqrt{1-\frac{v^2}{c^2}}\>.
$
This tells us that, given $t'_{i-1}=t'_i$, $t_i$ will be larger than
$t_{i-1}$ by $\gamma \frac{v}{c^2} (x'_i-x'_{i-1})$, the difference
being proportional to the length $x'_i-x'_{i-1}$ of the segment. All
that is important is that by repeating the argument for all segments,
we find that we must have $t_0<t_1<\ldots<t_{i}<\ldots<t_{N-1}<t_N$.
But point number 0 and point number $N$ are identical. Hence we have
$t_0<t_N=t_0$, a clear and evident contradiction.

\emph{Uniformly} accelerated rotation is possible, accelerated
\emph{rigid} rotation is not.

Note that a disk rotating at \emph{constant} angular velocity
\emph{does} conform with the definition of Born rigid motion, if we
assume Einstein's result to be correct. In fact, the notion of Lorentz
contraction refers to the comparison of lengths of the same object in
two different frames, not to that of lengths before and after
acceleration. \footnote{The length change on acceleration depends on
  the precise acceleration program and may include elastic
  deformations. Bell's spaceship paradox provides an example.
}  According to Einstein, the length of the circumference of
the rotating disk is $L= 2\pi R$ in the non-rotating frame and $L'=
\gamma L$ in the rotating one, so the circumference is reduced by
precisely the factor $1/\gamma$ in the frame with respect to which it
moves.\footnote{Of course, the right amount of Lorentz contraction
  would also be present, if the circumference was $2\pi R/\gamma$ in
  the non-rotating frame and $2\pi R$ in the rotating one. But then we
  would have the paradoxical situation described by Ehrenfest, unless
  we assumed space to be positively curved in an \emph{inertial} frame, for
  which there is no reason.}

\section{Realizability}

A typical argument against the discussed arrangement of two disks is
that the material properties of disks make it impossible to keep the
rotating disk flat or to keep the two disks on top of each other.  In
a discussion, a student once argued that the top disk would have to
spin so fast that it would fly away. The argument is invalid for the
simple reason that we do not need a disk that is spinning extremely
fast.  Relativity is not only valid at large speeds!  Its
predictions hold at any speed, we just have to measure more precisely
to detect and quantify effects at smaller speeds. 

Consider a steel disk of 1~m diameter and a mass of 1000~kg (this
corresponds to $\approx 16$ cm height). Since the transversal speed of
sound in steel is about 3200~m/s, we expect its elastic yield limit
not to be approached before parts of it are rotating at this speed. To
be safe, let us rotate it at 500 rps, meaning that its circumference
moves at about 1500~m/s, i.e., $v/c\approx 5 \times 10^{-6}$.  Special
relativistic effects are $\mathcal{O}\left((v/c)^2\right)$, so we need
a precision in time measurement of 1 part in $4\times10^{10}$.  With
current atomic clocks, precisions that are better than this by a
factor of $2.5\times 10^4$ are achievable. Chip-scale atomic clocks
were mounted at NIST already in 2004
(\emph{http://tf.nist.gov/ofm/smallclock/CSAC.html}), albeit with a
precision of 10$^{-10}$ only. This would not yet suffice, but is only
off by a factor of 4 (while big atomic clocks achieve 10$^{-15}$).
However, that happened several years ago and there is little doubt
that today ingenious experimenters would find a way to measure times
and lengths\footnote{Length measurements can be reduced to time
  measurements without loss of accuracy using light signals, because
  the speed of light is known exactly by definition.} on a rotating
disk with the required precision to observe non-Euclidean geometric
effects.

Another argument holds that general relativistic effects might mask or
even overwhelm special relativistic ones, due to the mass of the
disk.\footnote{Since general relativity encompasses special
  relativity, all special relativistic effects are strictly speaking
  also general relativistic ones. What is meant, is however
  clear: general relativistic effects as opposed to special
  relativistic ones are effects that are due to true gravitational
  sources, effects described by the field equations of general
  relativity that cannot be derived from special relativity.} These
effects arising from the stress-energy tensor of the system (and
including the effects of mechanical stresses in the disk) may be
simply estimated for our small-mass system, they are on the order of
$\Phi/c^2$, where $\Phi$ is the Newtonian gravitational potential. The
absolute value of the potential of our disk is certainly smaller than
that of a sphere of the same mass having as diameter the smallest
extension of the disk, i.e., 16 cm. We find
$\left\vert\Phi\right\vert\!/c^2< 10^{-23}$ which is below the discussed
measuring accuracy, so separability of the effects of acceleration and
true space-time curvature would not be an issue in our example.

Observers on the non-rotating disk are inertial. In inertial frames,
Euclidean geometry holds, so for these observers both disks will have
a circumference of $2\pi R$. Why does Lorentz contraction not reduce
the circumference of the rotating disk?  Of course, on setting the
disk in rotation, the material elements in its periphery will try to
Lorentz contract.  But they cannot do so arbitrarily as they are
connected to other material elements. What will happen instead is that
the rim of the disk will develop tensile tangential stresses. As
discussed in the last section, the situation is quite similar to the
one in Bell's spaceship paradox.

An alternative way of viewing this is to say that the material of the
disk does Lorentz contract but that it is in addition elastically
strained in a way that compensates the contraction. Incidentally, the
\emph{radius} of the disk will tend to shrink due to the tensile
tangential stresses in the disk, a tendency that may be partially
compensated or even overcompensated by centrifugal forces. So in order
to have two disks precisely of the same radius, the one that is
rotating may have to be taken of different size before starting to
spin it. It is also possible to make a stress-free rotating disk by
spinning a mould filled with molten steel and cooling it to perform
solidification while the whole arrangement is rotating. A similar
argument has been given by Rindler.\cite{rindler01} The resulting
steel disk will experience \emph{compressive} tangential stresses
along its rim when it is \emph{not} rotating anymore.

It should be noted that details of how the rotating disk might be
realized are irrelevant for an understanding of Einstein's thought
experiment.  All that is important is that we have a disk spinning at
constant angular velocity and a non-rotating disk that has the
same radius as the rotating one.

\section{Metric description}

In general relativity, the geometry of space-time is described by a
metric tensor $(g_{ik})$, generalizing the Minkowski metric
$(\eta_{ik}) = \text{diag}(-1,1,1,1)$. 
Moreover, the general relativistic metric may vary as a
function of space and time. The line element $\D s$ of
four-dimensional space-time is given by $\>\D s^2 = g_{ik} \D x^i \D
x^k\>,$ where Einstein's summation convention has been used (the
subscripts and superscripts running from 0 to 3). All local geometric
properties of space-time can be deduced from $(g_{ik})$.

Points on our disks may be specified in terms of cylindrical
coordinates, in which the Minkowski line element of the inertial
observer $C$ at the center is rewritten as
\begin{equation}
\D s^2 = - c^2 \D t^2 + \D r^2 + r^2 \D \varphi^2 + \D z^2\>.
\label{eq:minkow_cyl}
\end{equation}
We might suppress the summand $\D z^2$, as $\D z=0$ on the disks, but
we keep it to remind ourselves of the four-dimensional nature of
space-time.  The non-rotating disk would then be given by, say, $0\le
r\le R$, $0\le \varphi< 2\pi$, $z=-\Delta z$, and $t$ varies through
its lifetime.\footnote{For more realism, we might give the disk a
  finite thickness, letting $z$ vary between $-\Delta z_1$ and
  $-\Delta z_2$.} Each material point on the disk would have fixed
coordinates $r$, $\varphi$, and $z$. The rotating disk would be
described in terms of the same coordinates by $0\le r\le R$, $0\le
\varphi< 2\pi$, $z=0$
%\footnote{Or $0 \le z \le \Delta z_0$.} 
and $t$ varying through its lifetime. Its material points would have
$r$ and $z$ fixed and $\varphi = \omega t+\varphi_p$ mod $2\pi$, with
$\omega$ the angular velocity of the disk, assumed constant, and
$\varphi_p$ the angular position of the %considered
point at time $t=0$. The rim velocity is $v=\omega R$ and of course
$v<c$, i.e., for given $\omega$, $R$ cannot be larger than $c/\omega$.
% , whereas for given $R$, $\omega$ has the upper bound $c/R$).

Now we wish to describe the rotating disk from the point of view of
comoving observers, meaning that material points have fixed
coordinates.  An important question then is how to choose the
\emph{time} coordinate -- after all, the proper times of observers
sitting at different radial coordinates of the disk run at different
rates due to different time dilation factors.  As it turns out, there
is no way to use this local proper time as a global time coordinate on
the disk for any useful time interval.\footnote{One can formally
  rewrite the metric in terms of the local proper time. Then
   the prefactor of $\D r^2$ in the line element becomes time
  dependent and negative as time increases. We might put up with a
  time dependent metric, but not with one where the spatial part has
  the wrong signature.}  However, one of the nicer features of general
relativity is that the choice of time coordinate is (with minor
restrictions) as arbitrary as the choice of spatial coordinates. This
freedom of choice comes at a price -- velocities change on a
redefinition of time, so the postulate of the constancy of the speed
of light does not hold for coordinate velocities.  Light propagation
is instead described by $ds^2=0$.
% As we shall see, there is nevertheless a sense in which we can still
% maintain that the vacuum speed of light is a universal constant.

A convenient choice of the time coordinate is to just keep the time
$t$ of the central observer $C$.
% In principle, one might even construct clocks that reproduce this
% time on the disk. All clocks at a fixed distance $r$ from the center
% of the disk should be initialized (and resynchronized when
% necessary) by a light pulse emitted from the center. In addition,
% they should be set up in a GPS like fashion to run fast by a factor
% of $1/\sqrt{1-\omega^2 r^2/c^2}$ with respect to their proper time.
So we take as new coordinates
\begin{equation}
  \tpp = t\>, \quad \rpp = r\>,\quad \ppp=\varphi-\omega t\>\>\text{mod }2\pi\>,
  \quad \zpp=z 
\label{eq:disk_coord}
\end{equation}
which transforms the line element \eqref{eq:minkow_cyl} into
\begin{align}
\D s^2 = &-\left(1-\frac{\omega^2 \rpp^2}{c^2}\right) c^2 \D \tpp^2  
+ 2 \frac{\omega\rpp^2}{c} \,c\D \tpp \,\D \ppp
\nonumber\\ 
&+\D \rpp^2 
+ \rpp^2 \D \ppp^2 + \D \zpp^2\>.
\label{eq:disk_line_el}
\end{align}
The proper time interval $\D\tau$ for an observer sitting at a point
with fixed $\rpp$, $\ppp$, and $\zpp$ can be read off this equation
using $\D s^2=-c^2\D\tau^2$ and setting $\D\rpp=\D\ppp=\D\zpp=0$:
\begin{equation}
\D\tau = \sqrt{1-\frac{\omega^2 \rpp^2}{c^2}}\D\tpp\>,
\label{eq:proper_time}
\end{equation}
as expected from time dilation for a velocity $\omega\rpp$.

At this point, two things may be noted. First, the metric describes a
frame that is not time-orthogonal.  The components of the metric
tensor are scalar products of base vectors spanning the four-space
under consideration. Therefore, if the metric contains nondiagonal
terms (here summing up to the pre\-factor of the $\D\tpp\,\D\ppp$
term), the corresponding pair of base vectors is not orthogonal. In
our case, we have a mixing of the time and angular coordinates.
 Second, the metric
\eqref{eq:disk_line_el} is obtained from the Minkowski metric
\eqref{eq:minkow_cyl} by a mere coordinate transformation. Hence, it
cannot describe a space-time with a different geometry. Coordinates
are just a way to label space-time points, they do not modify nor
create the geometry. This means that the space-time of the rotating
disk is just the Minkowski space-time, it is flat, there is no
curvature.\footnote{Therefore, we would not really need general
  relativity to describe space-time aspects of the rotating disk.} The
situation changes when one is dealing with true gravitating bodies --
they curve space-time.

That space-time is flat does not mean that space has to be flat, as
can be easily seen from appropriate analogs in Euclidean space. The
surface of a sphere is clearly curved, it has nonvanishing Gaussian
curvature. Still the sphere is part of a flat Euclidean space, so a
flat space can accommodate a curved space of smaller dimensionality.
Just as the surface of a sphere may be obtained by setting one of the
spherical coordinates of three-space, the radial one, equal to a
constant, we may construct a hypersurface of four-space by setting the
time variable equal to a constant -- this hypersurface will correspond
to a spatial slice of four-space. Doing so in
\eqref{eq:disk_line_el}, we obtain the spatial line element given by
\begin{equation}
\D\ell_C^2 = \D\rpp^2 + \rpp^2 \D\ppp^2 + \D\zpp^2\>,
\label{eq:euclid_line} 
\end{equation}
which is the standard form of the line element of Euclidean geometry
in cylindrical coordinates. Hence, the three-geometry defined by the
hyperplane of simultaneity $\tpp=\text{const.}$ is flat. This is not
 too surprising, given that the time coordinate $\tpp$ is identical
to $t$, the time of an inertial observer ($C$).

However, the splitting of space-time into $(1+3)$-di\-mens\-ional
submanifolds describing time and space is not unique, so the resulting
three-dimensional geometry depends on the chosen time coordinate,
which corresponds to a particular synchronization procedure.  The time
coordinate $\tpp$ does not conform with Einstein
synchronization\footnote{To Einstein synchronize clock $B$ with clock
  $A$, send a light signal from $B$ to $A$ to be returned immediately
  with the clock reading of $A$. Then set the time on $B$ to the time
  reading from $A$ plus half the time span elapsed on $B$ between
  emission and reception of the signal.} for the local clocks of disk
observers. This is obvious for clocks at different radii corresponding
to different rates of proper time, but it is also true for clocks
having the same radial coordinate, as we shall see below.  In general,
there is a great variety of possible synchronization choices,
essential restrictions being only that the simultaneity relation so
established be symmetric and transitive and that simultaneous events remain
space-like. Any space-like foliation of space-time (if one exists) may
serve to define three-dimensional sets of simultaneous events and thus
give a prescription for the synchronization of (non-standard) clocks.
For example, it is quite permissible to define a new time coordinate
on the disk via $\tbb = \tpp- a\rpp\>, $ where $a$ is some constant
satisfying $a<1/c$.  Rewriting the metric in terms of this new time
coordinate and the old spatial coordinates (i.e., letting $\rbb=\rpp$,
$\pbb=\ppp$, $\zbb=\zpp$) and setting $\D \tbb=0$, we obtain for the
spatial line element
\begin{align*}
\D\lbb^{\,2} &= \left[ 1-a^2\left( c^2 - \omega^2\rbb^2
\right)
\right]\D\rbb^2+ 2\omega\rbb^2 a \,\D\rbb \,\D\pbb
\\ 
&\hspace*{2mm}+\rbb^2 \D\pbb^2 +
\D\zbb^2\>.
\end{align*}
To assess what kind of geometry is described by this spatial metric,
we calculate the measured radius $\bar{R}$ and circumference $\bar{L}$
of a circle with coordinate radius $R$ %(setting $\D\pbb=\D\zbb =0$)
 %(setting $\D\rbb=\D\zbb =0$)
\begin{align*}\bar{R}&= \bskipdm\int\limits_{\D\pbb=\D\zbb =0}
 \bskipdm\D\lbb =
  \int_0^R \sqrt{\rule{0mm}{2.8mm} 1-a^2\left( c^2 -
      \omega^2\rbb^2\right)}\,\D\rbb < R \>,
  \\
  \bar{L} &= \bskipdm\int\limits_{\D\rbb=\D\zbb =0} \bskipdm\D\lbb =
  \int_0^{2\pi} R \D\pbb = 2\pi R\>,
\end{align*}
from which we conclude that
$\bar{L}/\bar{R}>2\pi$, hence the spatial geometry connected with this
global time coordinate is non-Euclidean and hyperbolic. Note that the
time coordinate $\tbb$ has no particular significance, it was just
introduced to demonstrate that the spatial geometry may depend on the
choice of clock synchronization.

Klauber\cite{klauber98} insists that ``regardless of how one believes
time should be defined on the disk'' the geometry of the latter must
be Riemann flat, because any finite object traveling a geodesic path
in the plane of the disk surface will not experience any tidal
stresses. From general relativity, we expect objects moving along
geodesics (of space-time) to experience tidal forces, if and only if
\emph{space-time} is curved. Therefore Klauber's observation implies
that space-time must be flat. This is what we assumed all along. The
rotating disk is not supposed to have a sufficiently large mass for
gravity to play a role.
% In any case, we have not considered that role.
But that does not tell us anything about \emph{space}. In fact,
general relativity informs us that objects \emph{moving along
  geodesics} will not experience any tidal stresses in a flat
space-time, \emph{regardless} of how it is divided into $(1+3)$
submanifolds.  \emph{These} tidal forces will not care at all about
spatial curvature!\footnote{However, if an extended body is
  \emph{not} allowed to move along a geodesic of space-time, then it
  may experience tidal forces. An object kept at a fixed radius $\rpp$
  will be subject to centrifugal forces  known to produce
  tidal stresses.}

Neither this hyperbolic geometry nor the aforementioned Euclidean one
correspond to what an observer at rest on the rotating disk will
measure using standard clocks and standard rulers. Standard clocks
run at the rate of her proper time and the proper length of a
standard ruler is given by $c/2$ times the time a light signal will
take on a standard clock (which is at rest with respect to the ruler)
to travel from one end of the ruler to the other and
back.\footnote{Defining the spatial line element by a slice of
  constant $t$ in a non-orthogonal frame, we give up using standard
  rulers for length definition. Introducing $\tbb$ on the
  \emph{non-rotating} disk would also render its geometry hyperbolic.}

How can we obtain the geometry on the rotating disk as measured by
standard rulers?  We may employ one of the pillars on which general
relativity is founded, the equivalence principle. It states that in a
sufficiently small freely falling system of reference the laws of
physics have the same form as in the inertial frames of special
relativity. Expressed differently, local freely falling frames
\emph{are} inertial frames.  In mathematical terms, this means that
for any metric describing a piece of space-time (and the gravitational
fields arising from its curvature) there are coordinate
transformations mapping it locally to a Minkowski metric.  The laws of
physics in these small space-time patches are known from special
relativity. Transforming back to the original metric, we get the form
of physical laws in the presence of space-time curvature, i.e.,
gravitational fields. Transforming \emph{to} the Minkowski metric
locally, we obtain a local decomposition of space-time into the proper
time of the freely falling observer and a small platform of
three-space, in which standard rulers can be established just as in
special relativity.

Since the rotating disk is no inertial system, there is no
\emph{global} coordinate transformation mapping the global time to the
proper time of material points of the disk and establishing a common
spatial frame for them. However, to find a \emph{local}
transformation mapping the metric to Minkowskian form is easy. In
fact, since the squared line element is just a quadratic form, it can
be diagonalized by completion of squares, and a diagonal metric with
the correct signature can be reduced to the standard $(\eta_{ik})$ by
appropriate rescaling of the coordinate differentials,\footnote{For a
  local transformation, other than a global one, we do not have
  integrability conditions due to the fact that coordinate
  differentials should be total differentials.}
\begin{equation}
\begin{aligned}
  \D s^2 &= -\left(1-\frac{\omega^2 \rpp^2}{c^2}\right) \left(c \D
    \tpp - \frac{\omega \rpp^2}{c \left(1-\frac{\omega^2
          \rpp^2}{c^2}\right) }\D \ppp\right)^2
  \\
  &\hspace*{2mm}+\D \rpp^2 + \left(\rpp^2+ 
  \frac{\omega^2 \rpp^4}{c^2 \left(1-\frac{\omega^2
          \rpp^2}{c^2}\right)}\right)
 \D \ppp^2 + \D \zpp^2
 \\
&\equiv -c^2 \D\ttt^2 + \D \rtt^2 + \rtt^2 \D \ptt^2 + \D \ztt^2
\>,
\end{aligned}
\label{eq:diagonalized_metric}
\end{equation}
where 
\begin{equation}
\begin{aligned}
\D\ttt &= \sqrt{1-\frac{\omega^2 \rpp^2}{c^2}}\left(\D
    \tpp - \frac{\omega \rpp^2}{c^2 \left(1-\frac{\omega^2
          \rpp^2}{c^2}\right) }\D \ppp\right)\>,
 \\
\D\rtt&= \D\rpp\>,
 \\
\D\ptt&= \frac{1}{\sqrt{1-\frac{\omega^2 \rpp^2}{c^2}}}\D\ppp\>, 
\\
\D\ztt&= \D\zpp\>.
\end{aligned}
\label{eq:local_inertial_coords}
\end{equation}
In the last line of \eqref{eq:diagonalized_metric}, the metric has
Minkowskian form, i.e., its spatial part may be interpreted as
defining the \emph{proper} length element \footnote{There is an
  alternative way of determining the proper spatial line element via
  the calculation of the time a light signal will need to travel to a
  close-by point and back.\cite{landau59,gron75} While this is
  physically well-motivated, the calculation is slightly more
  complicated. %involving the solution of a quadratic equation.
  Moreover, its use of the mathematical statement of the equivalence
  principle is less direct.}
\begin{equation}
  \D\ell^2 = \D\rpp^2 + \frac{1}{1-\frac{\omega^2 \rpp^2}{c^2}} 
  \rpp^2\D\ppp^2 
  +\D\zpp^2\>.
\label{eq:proper_line}
\end{equation}
A circle with coordinate radius $\rpp=R$ will also have the measured
radius $R$, because for $\D\ppp=\D\zpp=0$ we have $\D\ell=\D\rpp$. Its
circumference will be
\begin{equation}
  \hat{L} = \bskipdm\int\limits_{\D\rpp=\D\zpp=0}\bskipdm \D\ell 
= \int_0^{2\pi} \frac{1}{\sqrt{1-\frac{\omega^2 R^2}{c^2}}} 
R  \D\ppp 
= \frac{2\pi R}{\sqrt{1-\frac{\omega^2 R^2}{c^2}}}\>.
\end{equation}
Since the speed of revolution of an observer $M$ at $\rpp=R$ about $C$
(at $\rpp=0$) is $v=\omega R$, this satisfies
\[ \hat{L}  = L' = 2\pi R \gamma > 2\pi R\>,     \qquad \gamma
= \frac1{\sqrt{1-\frac{v^2}{c^2}}}\>,
\]
which is the result obtained by Einstein\cite{einstein16} and
others\cite{berenda42,gron75,rizzi02}.  Hence, it seems expedient to
consider the hyperbolic geometry described by the proper line element
\eqref{eq:proper_line} the natural geometry of the rotating disk.
% \footnote{Which is, incidentally, also the geometry of the
%   non-rotating disk from the point of view of observers on the
%   rotating disk, although they will find direct measurements proving
%   this difficult to perform.}. 
It should be emphasized that this geometry is \emph{not} obtainable
from a hypersurface of constant time in any synchronization. Instead,
it may be visualized as constructed from local patches of space
orthogonal to the world lines of material points on the disk. Although
the construction to obtain the corresponding proper length element has
been known and used correctly by a number of authors
\cite{berenda42,gron75,landau59,moeller52,arzelies66}, it has been put
on a rigorous formal basis only recently via definition of the
so-called \emph{relative space}.\cite{rizzi02,ruggiero03} In this
approach, world lines are used to define equivalence classes
constituting the points and local space platforms of relative space,
which then becomes the quotient space of the world tube of the disk
referred to these classes. This is a precise formulation of our
introductory observation that space is only defined by a collection of
observers (which are test particles in Ref.~\onlinecite{rizzi02}).
The relative-space approach is claimed to be generalizable to
non-stationary and non-symmetric frames of reference,\cite{rizzi02}
whereas in preceding treatments of the rotating disk, the definition
of the union of space platforms as a globally valid geometry relied on
the fact that the metric is independent of time. The procedure of
laying out physical rulers is indefinitely repeatable (always giving
the same spatial geometry) only if the metric does not change with
time. In this case, it establishes a spatial geometry that is
independent of the notion of simultaneity, because the length of a
standard ruler at rest with respect to an observer can be ascertained
with a single clock without the necessity of synchronizing several
clocks.% \footnote{We always assume standard rulers to be sufficiently
%   short that there is no difference between their length measured as
%   radar distance\cite{rindler01} and measured as ruler distance
%   (breaking it up into the sum of lengths of smaller rulers).}

Let us now turn to the geometry of the non-rotating disk as assessed
by observers on the rotating disk. This is interesting, because the
non-rotating disk is a moving object according to these observers, and
it must somehow be embedded in their hyperbolic space. Observers on
the rotating disk at $\rpp=R$ may synchronize their standard clocks
according to the Einstein prescription, which means to use the time
$\ttt$.  Setting $\D\ttt=0$ in the first equation of
\eqref{eq:local_inertial_coords}, we have
\begin{equation}\D\tpp =\D t = \frac{\omega R^2}{c^2 \left(1-\frac{\omega^2
          R^2}{c^2}\right)}\D\ppp =  \frac{v R}{c^2} \gamma^2 \D\ppp
\label{eq:dtpp_dttt_0}
\end{equation}
and integrating from $\ppp=0$ to $\ppp=2\pi$ along the perimeter, we
find $\Delta t = 2\pi R \gamma^2 \frac{v}{c^2} = \gamma \frac{v}{c^2}
L'$.

To interpret this result let us imagine a standard clock $B$ that is
Einstein synchronized with some clock $A$ on the disk edge to be
taken around the disk instantaneously (i.e., it ticks off no time,
$\D\ttt=0$) in the positive angular direction and then be compared
with clock $A$.  Because according to the central observer, the time
$\Delta t$ has passed on the return of clock $B$ and clock $A$ is
running slow by a factor of $1/\gamma$ with respect to the central
clock, clock $A$ will have covered an interval $\Delta t'=\Delta
t/\gamma=\frac{v}{c^2} L'$ during $B$'s journey. Hence clock $B$ will
lag behind $A$ by $\Delta t'$.  In reality, clock $B$ cannot be moved
instantaneously, but if we move $B$ sufficiently slowly that time
dilation with respect to $A$ is negligible, then clock $B$ will have
ticked off an interval $\Delta t'$ less than $A$ when they meet again.
The phenomenon of the time gap $\Delta t'$ arising with this
synchronization method (which is closely related to the kinematic
resolution of Ehrenfest's paradox) has been discussed by many
authors,\cite{rizzi98,rizzi99,rizzi02,wucknitz04} with a particularly
transparent exposition given by Cranor et al.\cite{cranor99} As long
as we \emph{do not close} the curve along which we synchronize,
Einstein synchronization is however possible without contradiction.
 
Consider now the circumference of the non-rotating disk given by
$0\le\varphi<2\pi$, $r=R$. What angle will it cover on the rotating
disk? Since the non-rotating disk is moving with respect to the
spinning one, this question makes sense only with the proviso
``at a given time''. But then synchronization comes into play. If we
assume Einstein synchronization, we have to evaluate the angle at
$\D\ttt=0$, i.e., $\D\ppp $ is given by \eqref{eq:dtpp_dttt_0}. Moreover,
we have
\begin{equation}
  \D\varphi = \D\ppp+\omega \D t =  \D\ppp\left(1+\frac{v^2}{c^2}
    \gamma^2\right)
  = \gamma^2 \D\ppp\>,
\end{equation}
hence, when $\varphi$ runs from 0 to $2\pi$, $\ppp$ will only cover an
interval of length $2\pi/\gamma^2$. This is smaller than $2\pi$, so
Einstein synchronization was legitimate. Hence, we find that the
non-rotating disk does not fully cover the rotating one! The time gap
translates into a spatial gap, the circumference of the non-rotating
disk is not a closed curve for observers on the rotating disk. Of
course, the question immediately arises, what is \emph{in} the gap?
The answer is that there is no discontinuity. Since equal times for
corotating observers correspond to times increasing along the
direction of rotation for observers on the static disk, the
continuation of the latter is just another copy of itself but at later
times (and possibly more than one copy, depending on the value of
$\gamma$).  Discontinuities in the course of events on the
non-rotating disk will only seem to arise when observations are
compared at equal times on clocks that got out of synchronization,
being separated by a full turn around the rotating disk. But this is
easily explicable as a problem of those clocks, not one of the static
disk.

\begin{figure}[h!]
\includegraphics[height=3.0cm]{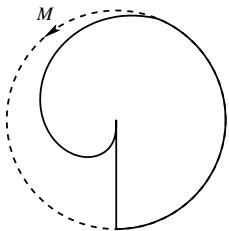}
\caption{Coverage of the rotating disk (dashed line) by the
  non-rotating one (solid line).  Projection from hyperbolic to
  Euclidean plane. $v(R)=0.75c$ $\Rightarrow$ $\gamma^2(v(R))=2.29$.}
\label{fig:nonrot_disk}
%\vspace*{-5mm}
\end{figure}

The filling of a spatial gap on the rotating disk by a later replica
of the non-rotating one is a consequence of the fact that the
static disk is really a space-time object and that pieces of it
belonging to different times are scrolled onto a single time for
observers on the rotating disk.  Its material points cover merely part
of the spinning disk, if counted only once, viz.~between two encounters of
a corotating observer with a (straight) line connecting the center of
the static disk with a point on its circumference. Space is nonetheless
hyperbolic and the missing part of the rotating disk is covered by an
additional copy (or more than one) of (part of) the non-rotating disk.
The shape of the first copy, obtained by measuring the ``material''
circumference at every radius along a line of Einstein simultaneity,
is shown in Fig.~\ref{fig:nonrot_disk}.

It should be pointed out that the observations of a traveler moving at
constant speed around a circle discussed by Semon et al.\cite{semon09}
are completely explicable by these phenomena without the need of
invoking elliptic geometry. Evidently, the situation considered is
equivalent to an observer $M$ on the rotating disk measuring the
length of the rim of the non-rotating disk.  A detailed description of
the measuring procedure is not given in Ref.~\onlinecite{semon09},
but it is obvious that the authors assume Einstein synchronization.
Then the relative speed of the rim of the non-rotating disk with
respect to the observer is $-v$.  The total time $M$ takes to complete
a full turn, as observed by $C$, around the circle is $\Delta t_0=
2\pi R/v$ according to $C$ and, taking time dilation into account,
$\Delta t'_0 = 2\pi R/v\gamma = L/v\gamma = L'/v\gamma^2\equiv L''/v$
according to $M$, who will conclude the length to be
$L''=L/\gamma <L$. This view can be defended, as we have seen.
However, the conclusion that this constitutes a measurement of the
circumference of a circle and that therefore the geometry is elliptic
fails, due to the fact that the curve is not even closed in $M$'s
reference frame, because the perimeter of the non-rotating disk covers
only $1/\gamma^2$ of the circumference of the disk, on which $M$ is
living.  Note that we \emph{do} have Lorentz contraction of the
non-rotating disk here: in its own frame its circumference is $L$, in
$M$'s frame (with respect to which it rotates) it is $L/\gamma$,
i.e., contracted -- but it is not a full circle.

The description of the non-rotating disk is remarkably complex, if
Einstein synchronized clocks are used on the spinning disk. It becomes
much simpler if we admit a different synchronization, discussed by
Cranor et al.\cite{cranor99} To realize it, send a light flash from
the center of the rotating disk and have all observers on its rim set
their standard clocks to the same time on arrival of the light signal.
Let us call this method central synchronization.  The relationship
between increments of ``local center-triggered time'' $\tvv$ and
``local Einstein time'' $\ttt$ is
\begin{equation}
\D\tvv = \D\ttt + \frac{\omega R^2}{c^2 \sqrt{1-\frac{\omega^2
          R^2}{c^2}}}\D \ppp\quad \Leftrightarrow \quad \D\tvv = \D\tpp/\gamma\>,
\label{eq:central_synch}
\end{equation}
i.e., clocks at fixed radius are advanced by a fixed amount of time
proportional to $\ppp$ with respect to Einstein synchronization. This
amount is just sufficient to close the time gap and to allow clocks to
be synchronized around the disk.\footnote{\emph{Standard} clocks at
  \emph{different} radii cannot be synchronized with each other this
  way.} The second equality of \eqref{eq:central_synch} shows that
centrally synchronized clocks share the notion of simultaneity with
the central observer $C$ ($\D\tvv=0$ implies $\D\tpp=0$ and vice
versa) and likewise with observers on the non-rotating disk. To
determine the length of the circumference of the static disk, $M$ may
still use the time $\Delta t'_0$ needed to circle around it -- this
interval depends only on the \emph{rate} of her clock, not on its
initial setting. But the relative velocity of fixed points on the rim
of the non-rotating disk with respect to $M$ is now different. To
calculate it, we set $\D\varphi=0$ ($\Rightarrow \D\ppp=-\omega\D t$)
at $r=\rpp=R$. This yields ($\D \pvv=\D \ptt$):
\begin{equation}
  v_c\equiv R\frac{\D \pvv}{\D \tvv} %= R \frac{\D \ptt}{\D \tvv} 
\underset{\eqref{eq:local_inertial_coords},\eqref{eq:central_synch}}{=}
 R\frac{\gamma\D\ppp}{\D\tpp/\gamma} = 
\gamma^2 R \frac{-\omega \D\tpp}{\D\tpp} = -\gamma^2 v\>.
\end{equation}
Hence, $M$ obtains as length of the material circumference of the
non-rotating disk $\Delta t'_0 \gamma^2 v = (L/\gamma v) \gamma^2 v =
\gamma L = L'$. The non-rotating disk now covers the rotating one
completely and has the same spatial geometry, its circumference is a
closed curve for $M$ and agrees with that of her own disk.

At this point, it may be appropriate to discuss another objection by
Klauber against the non-Euclidean nature of the space of the rotating
disk. He considers a tape measure around the rim of the
disk\cite{klauber98} and states it not to ``meet up with itself at the
same point in time''. This is of course the time lag discussed in the
last few paragraphs and rendering richer the interpretation of
measurements of the \emph{non-rotating} disk -- which would be moving
with respect to the tape.  However, there is no problem for the
\emph{rotating} disk on which the tape is at rest. Klauber's argument
again neglects the difference between space and space-time. While a
tape measure may be considered an approximately one-dimensional object
in space, it is two-dimensional in space-time, and its world sheet
does meet up with itself, not at a point in time, but along a whole
world line. For the purpose of measuring the length of the edge, it
does not matter at all whether one end of the tape is in place a
little earlier than the other. All that matters is that the two marks
on the tape the readings of which constitute the act of measurement,
have a common piece of world line, along which they can be compared.

\section{Conclusions}
As we have seen, general relativity allows several spatial geometries
to be associated with the set of observers on the rotating disk. This
is due to the non-uniqueness of the splitting of space-time into space
and time, which is closely related to the fact that clock
synchronization is conventional\cite{rizzi04a}, i.e., can be largely
done at liberty (as long as contradictions such as the assigning of
two different times to the same space-time event are avoided).

Locally, this splitting can be made unique by specifying a world line
describing an observer defined to be at rest and an infinitesimal
space platform orthogonal to it. If moreover a \emph{set} of observers
whose world lines fill a finite piece of space-time can be regarded as
being at rest -- operationally this means that light signals sent by
one of these observers to another and reflected back always take the
same time\footnote{In practice this condition is satisfied, if the
  observers correspond to material points of a non-deforming solid.}
-- this leads to a geometry that is distinguished by not being
dependent on synchronization.  In the case of our disk, this is its proper
geometry, corresponding to a frame of reference that coincides with
its local rest frame everywhere.\cite{rizzi02}

When geometric dimensions of objects are being measured that move with
respect to this reference frame, then synchronization dependence is
inevitable as was exemplified by two possible descriptions of the
non-rotating disk in the frame of the spinning disk. However, no
matter what synchronization is chosen, these objects will ``fit'' into
the space defined by the rest frame. The static disk, as observed by
inhabitants of the rotating disk, sits in a locally hyperbolic space
whether we choose Einstein synchronization, where length measurement
of its circumference would suggest otherwise, or central
synchronization, where the measured length  matches well with
our conception about the type of non-Euclidean geometry. In
the first case, the non-rotating disk manages to fill hyperbolic space
due to its space-time nature, conveniently providing different-age
repetitions of itself as needed.  Observers on the rotating disk might
note this repetitive nature.

On the other hand, the spatial geometry of the non-rotating disk, as
measured by its own inhabitants using standard rulers is simply
Euclidean. And so is the geometry of the rotating disk, which is
embedded in this Euclidean space. We have not investigated how a
different choice of synchronization by static-disk observers would
affect the geometry of the rotating disk though we can say that no
matter how that geometry changes it has to fit into a Euclidean space,
and the geometry of the static disk itself will be unaffected.

Nikoli{\'c} approaches the problem of length measurements on the
rotating disk by asserting that each point on the disk has its own
reference frame.\cite{nikolic00} Such a view seems quite natural from
a special relativistic background. In special relativity, a single
observer suffices to define a frame of reference, if we add that the
observer is inertial.
% and provide him with a tetrad consisting of a vector parallel to his
% world line and three orthogonal vectors spanning space.
Of course, given a single inertial observer in flat space-time, we
at once have infinitely many comoving ones with a
well-defined state of motion at arbitrary distances, a feature that
gets lost in curved space-times.

In general relativity, a single (typically non-inertial) observer
defines a \emph{local} reference frame at best, because the
continuation of his local platform of space (given by a small
hypersurface of simultaneity) is not unique. Generally, a physical
reference frame is a time-like congruence, a set of world lines of
observers or test particles filling space-time densely.\cite{rizzi02}
This congruence then defines a single frame of reference for a whole
set of observers, contrary to the statement by Nikoli{\'c}.

Very often, frames of reference are not even specified in general
relativity but only coordinate systems. In terms of these, a frame of
reference may be considered an equivalence class of coordinate
systems\cite{moeller52,rizzi02}, connected by so-called
\emph{internal} transformations, where the new space coordinates may
depend only on the old ones but not on time. The general form of such
a coordinate transformation is\cite{rizzi02}
\begin{align*}
{x'}^0 &= {x'}^0\left(x^0,x^1,x^2,x^3\right)\>,
\quad
{x'}^i = {x'}^i\left(x^1,x^2,x^3\right)\>,
\end{align*}
with the side conditions $g_{00}> 0$, $\partial {x'}^0/\partial
{x}^0>0$ (making sure that both $x^0$ and ${x'}^0$ are time
coordinates) and $g_{ij} \D x^i \D x^j<0$ ($i,j = 1,\ldots 3$,
allowing the interpretation of $x^i$, $i=1\ldots3$ as spatial
coordinates). Due to these  conditions, not all coordinate
systems are admissible as representations of a reference frame. For
instance, in the Schwarzschild metric,\cite{rindler01} $g_{00}$
changes sign on crossing the event horizon. So it describes a
single reference frame only when restricted to the region outside the
horizon. Coordinate frames are thus both more general and less
general than physical reference frames. They are more general, because one
global coordinate system may comprise more than one frame of reference
-- the Schwarzschild metric may be used to describe an observer
falling across the horizon. They are less general, because the same
reference frame is describable by many coordinate systems.

As pointed out by Nikoli{\'c},\cite{nikolic00} the spatial line
element \eqref{eq:proper_line} cannot be used for radii
$\rpp\ge\Rom\equiv c/\omega$. But this is not required either, as it is
the line element of observers \emph{at rest on a rotating disk}, and
such a disk must have a radius smaller than $\Rom$.  While it may be
debatable that the rotating coordinates describe a \emph{reference frame}
extending through all of space ($g_{00}$ changes sign at $\rpp=\Rom$),
they 
%coordinates \eqref{eq:disk_coord} and the metric \eqref{eq:disk_line_el} 
are perfectly applicable beyond $\Rom$.  Language is often sloppy and
a clear distinction not made in general relativity between coordinate
systems and reference frames. Because of time-non-orthogonality, the
metric \eqref{eq:disk_line_el} does not even become singular at
$\rpp=\Rom$, its eigenvalues all remain nonzero. Moreover, $\Rom$
definitely does not have the properties of a horizon. What happens
beyond $\Rom$ is that the requirement $\D s^2<0$ for a time-like world
line cannot be satisfied with $\D\rpp=\D\ppp=\D\zpp=0$, meaning that
no observers can exist at rest.  Inertial frames are dragged in the
counterrotating direction.  Nikoli{\'c} mentions the stars moving at
superluminal velocity around the Earth with respect to the Earth's
rotating (coordinate) frame.  They \emph{have} to do so in order to
\emph{avoid} being faster than light with respect to a local inertial
system.  The motion of a star in the the equatorial plane is given by
$\ppp=-\omega \tpp +\varphi_0$, and null geodesics describing incoming
starlight satisfy
\begin{equation} \D s^2= 0\>\Rightarrow \>
  \left(\frac{\D\rpp}{\D\tpp}\right)^2\!+ \rpp^2
  \left(\frac{\D\ppp}{\D\tpp}\right)^2 \!+
  2\omega\rpp^2\frac{\D\ppp}{\D\tpp} = c^2-\omega^2 \rpp^2\>.
\label{eq:null_geodes}
\end{equation}
As long as $\rpp>\Rom$, we must have ${\D\ppp}/{\D\tpp}<0$, so there
are no constant-$\ppp$ light paths. In fact, \eqref{eq:null_geodes} is
solved by ${\D\rpp}/{\D\tpp}=-c$, ${\D\ppp}/{\D\tpp}=-\omega$, which
remains valid for $\rpp\le\Rom$. So the light from a star reaches us
along a spiral when viewed from the rotating coordinate frame, and one with
typically many turns ($4\times 365$ turns for a star at $\rpp=4$
lightyears). If the star is at $\rpp=R_s$, then its coordinate speed
is $\omega R_s$, which is larger than $c$ for all stars but the Sun.
The radar distance\cite{rindler01} of the star is $R_s/\gamma\approx
R_s$.  While radial ruler distances seem unproblematic, in general the
ruler distance is not well-defined beyond $\Rom$ in rotating
coordinates -- there is no way to lay out rods at rest with respect to
the coordinate frame. To extend length measurements via rulers, one
first would have to continue the definition of the physical reference
frame beyond $\Rom$ by specifying appropriately moving observers. The
definition of a reference frame given by Rizzi and
Ruggiero\cite{rizzi02} does not require the observers to be at rest
with respect to each other.  Nikoli{\'c}'s suggestion to use the line
element \eqref{eq:euclid_line} corresponds to taking the reference
frame of all observers at rest with respect to the inertial observer
$C$ and is a valid approach, but by no means the only option.

According to Nikoli{\'c}, there is something wrong with the standard
solution to Ehrenfest's paradox as he tries to show by the example of
a rotating ring (with radius $r'$) in a non-rotating circular gutter
(with radius $r=r'$).\cite{nikolic00}  He is bothered by the following problem.
In keeping with the prevalent view (which is also the one presented
here), the proper length of a ring rotating at speed $v$ would be
$2\pi r' \gamma(v)$.  The proper length of the gutter, in which the
ring turns, would be only $2\pi r=2\pi r'$. From the point of view of
the gutter, the ring is Lorentz contracted to a length of $2\pi r$, so
it fits inside the gutter.  But from the point of view of the ring,
the gutter should be Lorentz contracted, i.e., have a length $2\pi
r'/\gamma(v) < 2\pi r'\gamma(v)$. Yet, since the ring is inside the
gutter, the length of the gutter cannot be smaller than that of the
ring.

This problem has been completely resolved in the preceding section
without giving up the hyperbolic geometry of the space of the ring. As
often is the case in discussions of length contraction, the key is
that the problem has been (unintentionally) posed in terms that are
based on an absolute notion of simultaneity.

Let us unveil the relativistic solution in a somewhat pictorial way.
Imagine the rotating ring to be inhabited by observer ants, equipped
with clocks, who decide to measure the distance between white marks
they found at certain points on the gutter (in fact, there is only a
single white mark, but the ants do not know that yet).  All the ants
have to do is to line up at equal distances around the ring and to
find out which ants see a white mark at a predetermined time.  The
number of ants  between two successful observers (plus one) multiplied
by the length of an ant's watched segment is the (approximate) length
between the marks.
% If only a single ant sees the mark at the given time, all
% segments around the ring have to be counted and the length of the
% gutter is the same as the length of the ring.

Obviously, the result will depend on how the ants have synchronized
their clocks.  Suppose they use Einstein synchronization.  Then there
is necessarily a desynchronization gap $\Delta t'_0$ between the
clocks of one pair of ants, and they have to make sure to choose their
measuring instant such that this gap does not appear inside the length interval
to be determined. If there are $N$ ants, we can number them so that
the gap is between ant no.~1 and its immediate neighbour no.~$N$
(numbering is in the sense of rotation). Then the gap will not appear
between the two ants seeing the mark, if measurement is taken to be in
the direction of increasing ant number and the instant of
measurement is chosen to be slightly before the first white mark
passes ant no.~1 (the ants can estimate this time by observing several
revolutions before making the actual measurement). Let us further
assume that the gutter is slowly and uniformly (in its own frame)
heated and that all ants continually measure and log its local
temperature.  Temperature measurements of all ants will be compared
for the point in time when the length measurement is made.

What are the ants' observations? First, they note that the measured
temperature of the gutter is not uniform. It increases in the sense of
rotation, because what is the same time for the ants becomes
increasingly later for the gutter. Second, they find white marks at
distances $2\pi r'/\gamma^2$. That is, there are two white marks at
least, observed by two different ants, and they have different
temperatures.\footnote{If $\gamma$ is larger than $\sqrt{2}$, more
  than two ants will simultaneously see a white mark.}  While it is
not immediately obvious to the ants that this is actually the same
white mark, they may note, if the gutter is not perfectly regular, that
the scratch pattern near the white marks is always the same.  Hence,
they will conclude that the length of the gutter between two white
marks is smaller than the length of the ring but that there is a
suspiciously similar piece of gutter following, different only by a
higher temperature.

So the result is that the gutter manages to contain the ring although
it is shorter. This is due to the fact that length elements of the
gutter at a fixed ring time correspond to length elements at different
gutter times. Aligning them does not produce a closed curve in a
space-time diagram but a helix. If we could disregard that
Einstein synchronization of clocks around the full ring (and further)
leads to contradictions (we would have two ants at the same segment
with different clock readings but claiming to have synchronized
clocks), then the gutter might contain an object of arbitrary length!

The situation is far simpler, if the ants use central synchronization
for their clocks (employing either of the three methods described in
Ref.~\onlinecite{cranor99}). Then what is simultaneous for the gutter
is simultaneous for the ants. Hence, the gutter has the same
temperature everywhere at the moment of measurement. Moreover, only a
single ant will find the white mark in its segment. The length of the
gutter is $2\pi  r'\gamma$, i.e., identical to that of the ring. 

Note that certain well-known features of special relativity such as
mutual time dilation, mutual length contraction, reciprocity of
relative velocities (if inertial system $S'$ moves at velocity
$\boldsymbol{v}$ in $S$, $S$ will move at $-\boldsymbol{v}$ in $S'$)
and isotropy of inertial systems all depend on Einstein
synchronization. A detailed discussion of these aspects that do not
seem to be well-known is beyond the scope of this paper.  Einstein
synchronization is the most favored method of defining simultaneity in
inertial systems, among other reasons, because it leads to symmetry
between different inertial systems and makes them isotropic.  Also the
second postulate of special relativity is valid in this
synchronization as a statement on one-way velocities of light. With
arbitrary synchronizations, it remains true only referring to the
round-trip velocity of light along closed curves.\cite{rizzi04a}

Clearly, in a rotating system, different synchronizations may be more
favorable.  If we use central synchronization on the ring, there is no
Lorentz contraction of the gutter, but the ring is still Lorentz
contracted in the frame of the gutter.  So the mutuality of length
contraction gets lost.

\bibliographystyle{apsrev4-1long}

%\bibliography{relativity}

\begin{thebibliography}{10}%
\makeatletter
\providecommand \@ifxundefined [1]{%
 \ifx #1\undefined \expandafter \@firstoftwo
 \else \expandafter \@secondoftwo
\fi
}%
\providecommand \@ifnum [1]{%
 \ifnum #1\expandafter \@firstoftwo
 \else \expandafter \@secondoftwo
\fi
}%
\providecommand \enquote [1]{``#1''}%
\providecommand \bibnamefont  [1]{#1}%
\providecommand \bibfnamefont [1]{#1}%
\providecommand \citenamefont [1]{#1}%
\providecommand\href[0]{\@sanitize\@href}%
\providecommand\@href[1]{\endgroup\@@startlink{#1}\endgroup\@@href}%
\providecommand\@@href[1]{#1\@@endlink}%
\providecommand \@sanitize [0]{\begingroup\catcode`\&12\catcode`\#12\relax}%
\@ifxundefined \pdfoutput {\@firstoftwo}{%
 \@ifnum{\z@=\pdfoutput}{\@firstoftwo}{\@secondoftwo}%
}{%
 \providecommand\@@startlink[1]{\leavevmode}%
 \providecommand\@@endlink[0]{}%
}{%
 \providecommand\@@startlink[1]{%
  \leavevmode
  \pdfstartlink
   attr{/Border[0 0 1 ]/H/I/C[0 1 1]}%
   user{/Subtype/Link/A<</Type/Action/S/URI/URI(#1)>>}%
  \relax
 }%
 \providecommand\@@endlink[0]{\pdfendlink}%
}%
\providecommand \url  [0]{\begingroup\@sanitize \@url }%
\providecommand \@url [1]{\endgroup\@href {#1}{\urlprefix}}%
\providecommand \urlprefix [0]{URL }%
\providecommand \Eprint[0]{\href }%
\@ifxundefined \urlstyle {%
  \providecommand \doi [1]{doi:\discretionary{}{}{}#1}%
}{%
  \providecommand \doi [0]{doi:\discretionary{}{}{}\begingroup
  \urlstyle{rm}\Url }%
}%
\providecommand \doibase [0]{http://dx.doi.org/}%
\providecommand \Doi[1]{\href{\doibase#1}}%
\providecommand \bibAnnote [3]{%
  \BibitemShut{#1}%
  \begin{quotation}\noindent
    \textsc{Key:}\ #2\\\textsc{Annotation:}\ #3%
  \end{quotation}%
}%
\providecommand \bibAnnoteFile [2]{%
  \IfFileExists{#2}{\bibAnnote {#1} {#2} {\input{#2}}}{}%
}%
\providecommand \typeout [0]{\immediate \write \m@ne }%
\providecommand \selectlanguage [0]{\@gobble}%
\providecommand \bibinfo [0]{\@secondoftwo}%
\providecommand \bibfield [0]{\@secondoftwo}%
\providecommand \translation [1]{[#1]}%
\providecommand \BibitemOpen[0]{}%
\providecommand \bibitemStop [0]{}%
\providecommand \bibitemNoStop [0]{.\EOS\space}%
\providecommand \EOS [0]{\spacefactor3000\relax}%
\providecommand \BibitemShut [1]{\csname bibitem#1\endcsname}%
%</preamble>
\bibitem{einstein16}%
  \BibitemOpen
  \bibfield{author}{%
  \bibinfo {author} {\bibfnamefont{A.}~\bibnamefont{Einstein}},\ }%
  \bibfield{title}{%
  \enquote{\bibinfo {title} {{Die Grundlage der allgemeinen
  Rela\-tivi\-t\"ats\-theo\-rie}},}\ }%
  \bibfield{journal}{%
  \bibinfo {journal} {Ann. Phys.}\ }%
  \textbf{\bibinfo {volume} {354}},\ \bibinfo {pages} {769--822} (\bibinfo
  {year} {1916}),\ \bibinfo {note} {{E}nglish Translation in \emph{The
  Principle of Relativity} (Methuen, 1923, {r}eprinted by Dover Publications,
  New York, 1952), pp. 109 -- 164.}%
  \bibAnnoteFile{Stop}{einstein16}%
\bibitem{berenda42}%
  \BibitemOpen
  \bibfield{author}{%
  \bibinfo {author} {\bibfnamefont{C.~W.}\ \bibnamefont{Berenda}},\ }%
  \bibfield{title}{%
  \enquote{\bibinfo {title} {{The Problem of the Rotating Disk}},}\ }%
  \bibfield{journal}{%
  \bibinfo {journal} {Phys. Rev.}\ }%
  \textbf{\bibinfo {volume} {62}},\ \bibinfo {pages} {280--290} (\bibinfo
  {year} {1942})%
  \bibAnnoteFile{NoStop}{berenda42}%
\bibitem{gron75}%
  \BibitemOpen
  \bibfield{author}{%
  \bibinfo {author} {\bibfnamefont{{\O}.}~\bibnamefont{Gr{\o}n}},\ }%
  \bibfield{title}{%
  \enquote{\bibinfo {title} {{Relativistic description of a rotating disk}},}\
  }%
  \bibfield{journal}{%
  \bibinfo {journal} {Am. J. Phys.}\ }%
  \textbf{\bibinfo {volume} {43}},\ \bibinfo {pages} {869--876} (\bibinfo
  {year} {1975})%
  \bibAnnoteFile{NoStop}{gron75}%
\bibitem{weber97}%
  \BibitemOpen
  \bibfield{author}{%
  \bibinfo {author} {\bibfnamefont{T.~A.}\ \bibnamefont{Weber}},\ }%
  \bibfield{title}{%
  \enquote{\bibinfo {title} {{A note on rotating coordinates in relativity}},}\
  }%
  \bibfield{journal}{%
  \bibinfo {journal} {Am. J. Phys.}\ }%
  \textbf{\bibinfo {volume} {65}},\ \bibinfo {pages} {486--487} (\bibinfo
  {year} {1997})%
  \bibAnnoteFile{NoStop}{weber97}%
\bibitem{rizzi98}%
  \BibitemOpen
  \bibfield{author}{%
  \bibinfo {author} {\bibfnamefont{G.}~\bibnamefont{Rizzi}}\ and\ \bibinfo
  {author} {\bibfnamefont{A.}~\bibnamefont{Tartaglia}},\ }%
  \bibfield{title}{%
  \enquote{\bibinfo {title} {{Speed of Light on Rotating Platforms}},}\ }%
  \bibfield{journal}{%
  \bibinfo {journal} {Found. Phys.}\ }%
  \textbf{\bibinfo {volume} {28}},\ \bibinfo {pages} {1663--1683} (\bibinfo
  {year} {1998})%
  \bibAnnoteFile{NoStop}{rizzi98}%
\bibitem{rizzi99}%
  \BibitemOpen
  \bibfield{author}{%
  \bibinfo {author} {\bibfnamefont{G.}~\bibnamefont{Rizzi}}\ and\ \bibinfo
  {author} {\bibfnamefont{A.}~\bibnamefont{Tartaglia}},\ }%
  \bibfield{title}{%
  \enquote{\bibinfo {title} {{On local and global measurements of the speed of
  light on rotating platforms}},}\ }%
  \bibfield{journal}{%
  \bibinfo {journal} {Found. Phys. Lett.}\ }%
  \textbf{\bibinfo {volume} {12}},\ \bibinfo {pages} {179--186} (\bibinfo
  {year} {1999})%
  \bibAnnoteFile{NoStop}{rizzi99}%
\bibitem{rizzi02}%
  \BibitemOpen
  \bibfield{author}{%
  \bibinfo {author} {\bibfnamefont{G.}~\bibnamefont{Rizzi}}\ and\ \bibinfo
  {author} {\bibfnamefont{M.~L.}\ \bibnamefont{Ruggiero}},\ }%
  \bibfield{title}{%
  \enquote{\bibinfo {title} {{Space Geometry of Rotating Platforms: An
  Operational Approach}},}\ }%
  \bibfield{journal}{%
  \bibinfo {journal} {Found. Phys.}\ }%
  \textbf{\bibinfo {volume} {32}},\ \bibinfo {pages} {1525--1556} (\bibinfo
  {year} {2002})%
  \bibAnnoteFile{NoStop}{rizzi02}%
\bibitem{ruggiero03}%
  \BibitemOpen
  \bibfield{author}{%
  \bibinfo {author} {\bibfnamefont{M.}~\bibnamefont{Ruggiero}},\ }%
  \bibfield{title}{%
  \enquote{\bibinfo {title} {{Relative space: space measurements on a rotating
  platform}},}\ }%
  \bibfield{journal}{%
  \bibinfo {journal} {Eur. J. Phys.}\ }%
  \textbf{\bibinfo {volume} {24}},\ \bibinfo {pages} {563--573} (\bibinfo
  {year} {2003})%
  \bibAnnoteFile{NoStop}{ruggiero03}%
\bibitem{pellegrini95}%
  \BibitemOpen
  \bibfield{author}{%
  \bibinfo {author} {\bibfnamefont{G.~N.}\ \bibnamefont{Pellegrini}}\ and\
  \bibinfo {author} {\bibfnamefont{A.}~\bibnamefont{Swift}},\ }%
  \bibfield{title}{%
  \enquote{\bibinfo {title} {{Maxwell's equations in a rotating medium: is
  there a problem?}}.}\ }%
  \bibfield{journal}{%
  \bibinfo {journal} {Am. J. Phys.}\ }%
  \textbf{\bibinfo {volume} {63}},\ \bibinfo {pages} {694--705} (\bibinfo
  {year} {1995})%
  \bibAnnoteFile{NoStop}{pellegrini95}%
\bibitem{weber97b}%
  \BibitemOpen
  \bibfield{author}{%
  \bibinfo {author} {\bibfnamefont{T.~A.}\ \bibnamefont{Weber}},\ }%
  \bibfield{title}{%
  \enquote{\bibinfo {title} {{Measurements on a rotating frame in relativity,
  and the Wilson and Wilson experiment}},}\ }%
  \bibfield{journal}{%
  \bibinfo {journal} {Am. J. Phys.}\ }%
  \textbf{\bibinfo {volume} {65}},\ \bibinfo {pages} {946--953} (\bibinfo
  {year} {1997})%
  \bibAnnoteFile{NoStop}{weber97b}%
\bibitem{klauber98}%
  \BibitemOpen
  \bibfield{author}{%
  \bibinfo {author} {\bibfnamefont{R.~D.}\ \bibnamefont{Klauber}},\ }%
  \bibfield{title}{%
  \enquote{\bibinfo {title} {{Comments regarding recent articles on
  relativistically rotating frames}},}\ }%
  \bibfield{journal}{%
  \bibinfo {journal} {Am. J. Phys.}\ }%
  \textbf{\bibinfo {volume} {67}},\ \bibinfo {pages} {158--159} (\bibinfo
  {year} {1998})%
  \bibAnnoteFile{NoStop}{klauber98}%
\bibitem{weber98}%
  \BibitemOpen
  \bibfield{author}{%
  \bibinfo {author} {\bibfnamefont{T.~A.}\ \bibnamefont{Weber}},\ }%
  \bibfield{title}{%
  \enquote{\bibinfo {title} {{Response to ``Comments regarding recent articles
  on relativistically rotating frames''}},}\ }%
  \bibfield{journal}{%
  \bibinfo {journal} {Am. J. Phys.}\ }%
  \textbf{\bibinfo {volume} {67}},\ \bibinfo {pages} {159--161} (\bibinfo
  {year} {1998})%
  \bibAnnoteFile{NoStop}{weber98}%
\bibitem{tartaglia99}%
  \BibitemOpen
  \bibfield{author}{%
  \bibinfo {author} {\bibfnamefont{A.}~\bibnamefont{Tartaglia}},\ }%
  \bibfield{title}{%
  \enquote{\bibinfo {title} {{Lengths on rotating platforms}},}\ }%
  \bibfield{journal}{%
  \bibinfo {journal} {Found. Phys. Lett.}\ }%
  \textbf{\bibinfo {volume} {12}},\ \bibinfo {pages} {17--28} (\bibinfo {year}
  {1999})%
  \bibAnnoteFile{NoStop}{tartaglia99}%
\bibitem{nikolic00}%
  \BibitemOpen
  \bibfield{author}{%
  \bibinfo {author} {\bibfnamefont{H.}~\bibnamefont{Nikoli{\'c}}},\ }%
  \bibfield{title}{%
  \enquote{\bibinfo {title} {{Relativistic contraction and related effects in
  noninertial frames}},}\ }%
  \bibfield{journal}{%
  \bibinfo {journal} {Phys. Rev. A}\ }%
  \textbf{\bibinfo {volume} {61}},\ \bibinfo {pages} {0321091--8} (\bibinfo
  {year} {2000})%
  \bibAnnoteFile{NoStop}{nikolic00}%
\bibitem{west08}%
  \BibitemOpen
  \bibfield{author}{%
  \bibinfo {author} {\bibfnamefont{J.}~\bibnamefont{West}},\ }%
  \bibfield{title}{%
  \enquote{\bibinfo {title} {{A relativistic rotating frame with physical
  majors, photons and mirrors: causality lost}},}\ }%
  \bibfield{journal}{%
  \bibinfo {journal} {Eur. J. Phys.}\ }%
  \textbf{\bibinfo {volume} {29}},\ \bibinfo {pages} {885--900} (\bibinfo
  {year} {2008})%
  \bibAnnoteFile{NoStop}{west08}%
\bibitem{semon09}%
  \BibitemOpen
  \bibfield{author}{%
  \bibinfo {author} {\bibfnamefont{M.~D.}\ \bibnamefont{Semon}}, \bibinfo
  {author} {\bibfnamefont{S.}~\bibnamefont{Malin}},\ and\ \bibinfo {author}
  {\bibfnamefont{S.}~\bibnamefont{Wortel}},\ }%
  \bibfield{title}{%
  \enquote{\bibinfo {title} {{Exploring the transition from special to general
  relativity}},}\ }%
  \bibfield{journal}{%
  \bibinfo {journal} {Am. J. Phys.}\ }%
  \textbf{\bibinfo {volume} {77}},\ \bibinfo {pages} {434--437} (\bibinfo
  {year} {2009})%
  \bibAnnoteFile{NoStop}{semon09}%
\bibitem{rizzi04}%
  \BibitemOpen
  \emph{\bibinfo {title} {{Relativity in rotating frames}}},\ edited by\
  \bibinfo {editor} {\bibfnamefont{G.}~\bibnamefont{Rizzi}}\ and\ \bibinfo
  {editor} {\bibfnamefont{M.~L.}\ \bibnamefont{Ruggiero}}\ (\bibinfo
  {publisher} {Kluwer Academic Publishers},\ \bibinfo {address} {Dordrecht},\
  \bibinfo {year} {2004})%
  \bibAnnoteFile{NoStop}{rizzi04}%
\bibitem{selleri96}%
  \BibitemOpen
  \bibfield{author}{%
  \bibinfo {author} {\bibfnamefont{F.}~\bibnamefont{Selleri}},\ }%
  \bibfield{title}{%
  \enquote{\bibinfo {title} {{Noninvariant One-Way Velocity of Light}},}\ }%
  \bibfield{journal}{%
  \bibinfo {journal} {Found. Phys.}\ }%
  \textbf{\bibinfo {volume} {26}},\ \bibinfo {pages} {641--664} (\bibinfo
  {year} {1996})%
  \bibAnnoteFile{NoStop}{selleri96}%
\bibitem{munoz10}%
  \BibitemOpen
  \bibfield{author}{%
  \bibinfo {author} {\bibfnamefont{G.}~\bibnamefont{Mu{{\~n}}oz}}\ and\
  \bibinfo {author} {\bibfnamefont{P.}~\bibnamefont{Jones}},\ }%
  \bibfield{title}{%
  \enquote{\bibinfo {title} {{The equivalence principle, uniformly accelerated
  frames, and the uniform gravitational field}},}\ }%
  \bibfield{journal}{%
  \bibinfo {journal} {Am. J. Phys.}\ }%
  \textbf{\bibinfo {volume} {78}},\ \bibinfo {pages} {377--383} (\bibinfo
  {year} {2010})%
  \bibAnnoteFile{NoStop}{munoz10}%
\bibitem{Note1}%
  \BibitemOpen
  \bibinfo {note} {Similarly, \protect \emph {any} observer at a fixed position
  on the rotating disk may be considered to belong to a set of instantaneously
  comoving inertial observers, living in Euclidean space, as well as to the set
  of stationary disk observers, whose space is non-Euclidean. Measurements take
  time, so the instantaneous inertial frame is not particularly practical for
  physical descriptions -- it changes continuously.}%
  \bibAnnoteFile{Stop}{Note1}%
\bibitem{wucknitz04}%
  \BibitemOpen
  \bibfield{author}{%
  \bibinfo {author} {\bibfnamefont{O.}~\bibnamefont{Wucknitz}},\ }%
  \enquote{\bibinfo {title} {{Sagnac effect, twin paradox and space-time
  topology -- Time and length in rotating systems and closed Minkowski
  space-times}},}\  (\bibinfo {year} {2004}),\
  \Eprint{http://arxiv.org/abs/gr-qc/0403111v1}{gr-qc/0403111v1}%
  \bibAnnoteFile{NoStop}{wucknitz04}%
\bibitem{dewan59}%
  \BibitemOpen
  \bibfield{author}{%
  \bibinfo {author} {\bibfnamefont{E.~M.}\ \bibnamefont{Dewan}}\ and\ \bibinfo
  {author} {\bibfnamefont{M.~J.}\ \bibnamefont{Beran}},\ }%
  \bibfield{title}{%
  \enquote{\bibinfo {title} {{Note on stress effects due to relativistic
  contraction}},}\ }%
  \bibfield{journal}{%
  \bibinfo {journal} {Am. J. Phys.}\ }%
  \textbf{\bibinfo {volume} {27}},\ \bibinfo {pages} {517--518} (\bibinfo
  {year} {1959})%
  \bibAnnoteFile{NoStop}{dewan59}%
\bibitem{bell76}%
  \BibitemOpen
  \bibfield{author}{%
  \bibinfo {author} {\bibfnamefont{J.~S.}\ \bibnamefont{Bell}},\ }%
  \enquote{\bibinfo {title} {{How to teach special relativity}},}\ in\
  \emph{\bibinfo {booktitle} {{Speakable and unspeakable in quantum
  mechanics}}}\ (\bibinfo {publisher} {Cambridge University Press},\ \bibinfo
  {year} {1993})\ pp.\ \bibinfo {pages} {67--80},\ \bibinfo {note} {first
  appeared in {Progress in Scientific Culture \textbf{1}} (1976)}%
  \bibAnnoteFile{NoStop}{bell76}%
\bibitem{gron81}%
  \BibitemOpen
  \bibfield{author}{%
  \bibinfo {author} {\bibfnamefont{{\O}.}~\bibnamefont{Gr{\o}n}},\ }%
  \bibfield{title}{%
  \enquote{\bibinfo {title} {{Covariant formulation of Hooke's law}},}\ }%
  \bibfield{journal}{%
  \bibinfo {journal} {Am. J. Phys.}\ }%
  \textbf{\bibinfo {volume} {49}},\ \bibinfo {pages} {28--30} (\bibinfo {year}
  {1981})%
  \bibAnnoteFile{NoStop}{gron81}%
\bibitem{ehrenfest09}%
  \BibitemOpen
  \bibfield{author}{%
  \bibinfo {author} {\bibfnamefont{P.}~\bibnamefont{Ehrenfest}},\ }%
  \bibfield{title}{%
  \enquote{\bibinfo {title} {{Gleichf\"ormige Rotation starrer K\"or\-per und
  Rela\-tivi\-t\"ats\-theorie}},}\ }%
  \bibfield{journal}{%
  \bibinfo {journal} {Phys. Zeitschrift}\ }%
  \textbf{\bibinfo {volume} {10}},\ \bibinfo {pages} {919--919} (\bibinfo
  {year} {1909})%
  \bibAnnoteFile{NoStop}{ehrenfest09}%
\bibitem{born09}%
  \BibitemOpen
  \bibfield{author}{%
  \bibinfo {author} {\bibfnamefont{M.}~\bibnamefont{Born}},\ }%
  \bibfield{title}{%
  \enquote{\bibinfo {title} {{Die Theorie des starren Elektrons in der
  Kinematik des Rela\-tivi\-t\"ats\-prin\-zips}},}\ }%
  \bibfield{journal}{%
  \bibinfo {journal} {Ann. Phys.}\ }%
  \textbf{\bibinfo {volume} {335}},\ \bibinfo {pages} {1--56} (\bibinfo {year}
  {1909})%
  \bibAnnoteFile{NoStop}{born09}%
\bibitem{rindler01}%
  \BibitemOpen
  \bibfield{author}{%
  \bibinfo {author} {\bibfnamefont{W.}~\bibnamefont{Rindler}},\ }%
  \emph{\bibinfo {title} {{Relativity. Special, general, and cosmological}}}\
  (\bibinfo {publisher} {Oxford Univ. Press},\ \bibinfo {address} {New York},\
  \bibinfo {year} {2001})%
  \bibAnnoteFile{NoStop}{rindler01}%
\bibitem{Note2}%
  \BibitemOpen
  \bibinfo {note} {The length change on acceleration depends on the precise
  acceleration program and may include elastic deformations. Bell's spaceship
  paradox provides an example.}%
  \bibAnnoteFile{Stop}{Note2}%
\bibitem{Note3}%
  \BibitemOpen
  \bibinfo {note} {Of course, the right amount of Lorentz contraction would
  also be present, if the circumference was $2\pi R/\gamma $ in the
  non-rotating frame and $2\pi R$ in the rotating one. But then we would have
  the paradoxical situation described by Ehrenfest, unless we assumed space to
  be positively curved in an \protect \emph {inertial} frame, for which there
  is no reason.}%
  \bibAnnoteFile{Stop}{Note3}%
\bibitem{Note4}%
  \BibitemOpen
  \bibinfo {note} {Length measurements can be reduced to time measurements
  without loss of accuracy using light signals, because the speed of light is
  known exactly by definition.}%
  \bibAnnoteFile{Stop}{Note4}%
\bibitem{Note5}%
  \BibitemOpen
  \bibinfo {note} {Since general relativity encompasses special relativity, all
  special relativistic effects are strictly speaking also general relativistic
  ones. What is meant, is however clear: general relativistic effects as
  opposed to special relativistic ones are effects that are due to true
  gravitational sources, effects described by the field equations of general
  relativity that cannot be derived from special relativity.}%
  \bibAnnoteFile{Stop}{Note5}%
\bibitem{Note6}%
  \BibitemOpen
  \bibinfo {note} {For more realism, we might give the disk a finite thickness,
  letting $z$ vary between $-\Delta z_1$ and $-\Delta z_2$.}%
  \bibAnnoteFile{Stop}{Note6}%
\bibitem{Note7}%
  \BibitemOpen
  \bibinfo {note} {One can formally rewrite the metric in terms of the local
  proper time. Then the prefactor of $\protect \mathrm {d}r^2$ in the line
  element becomes time dependent and negative as time increases. We might put
  up with a time dependent metric, but not with one where the spatial part has
  the wrong signature.}%
  \bibAnnoteFile{Stop}{Note7}%
\bibitem{Note8}%
  \BibitemOpen
  \bibinfo {note} {Therefore, we would not really need general relativity to
  describe space-time aspects of the rotating disk.}%
  \bibAnnoteFile{Stop}{Note8}%
\bibitem{Note9}%
  \BibitemOpen
  \bibinfo {note} {To Einstein synchronize clock $B$ with clock $A$, send a
  light signal from $B$ to $A$ to be returned immediately with the clock
  reading of $A$. Then set the time on $B$ to the time reading from $A$ plus
  half the time span elapsed on $B$ between emission and reception of the
  signal.}%
  \bibAnnoteFile{Stop}{Note9}%
\bibitem{Note10}%
  \BibitemOpen
  \bibinfo {note} {However, if an extended body is \protect \emph {not} allowed
  to move along a geodesic of space-time, then it may experience tidal forces.
  An object kept at a fixed radius $\protect \mathaccentV {hat}05E{r}$ will be
  subject to centrifugal forces known to produce tidal stresses.}%
  \bibAnnoteFile{Stop}{Note10}%
\bibitem{Note11}%
  \BibitemOpen
  \bibinfo {note} {Defining the spatial line element by a slice of constant $t$
  in a non-orthogonal frame, we give up using standard rulers for length
  definition. Introducing $\protect \mathaccentV {bar}016{t}$ on the \protect
  \emph {non-rotating} disk would also render its geometry hyperbolic.}%
  \bibAnnoteFile{Stop}{Note11}%
\bibitem{Note12}%
  \BibitemOpen
  \bibinfo {note} {For a local transformation, other than a global one, we do
  not have integrability conditions due to the fact that coordinate
  differentials should be total differentials.}%
  \bibAnnoteFile{Stop}{Note12}%
\bibitem{Note13}%
  \BibitemOpen
  \bibinfo {note} {There is an alternative way of determining the proper
  spatial line element via the calculation of the time a light signal will need
  to travel to a close-by point and back.\cite {landau59,gron75} While this is
  physically well-motivated, the calculation is slightly more complicated.
  Moreover, its use of the mathematical statement of the equivalence principle
  is less direct.}%
  \bibAnnoteFile{Stop}{Note13}%
\bibitem{landau59}%
  \BibitemOpen
  \bibfield{author}{%
  \bibinfo {author} {\bibfnamefont{L.~D.}\ \bibnamefont{Landau}}\ and\ \bibinfo
  {author} {\bibfnamefont{E.~M.}\ \bibnamefont{Lifshitz}},\ }%
  \emph{\bibinfo {title} {{The classical theory of fields}}},\ \bibinfo
  {edition} {{2$^\text{nd}$}}\ ed.\ (\bibinfo {publisher} {Pergamon Press},\
  \bibinfo {address} {Oxford},\ \bibinfo {year} {1959})%
  \bibAnnoteFile{NoStop}{landau59}%
\bibitem{moeller52}%
  \BibitemOpen
  \bibfield{author}{%
  \bibinfo {author} {\bibfnamefont{C.}~\bibnamefont{M{\o}ller}},\ }%
  \emph{\bibinfo {title} {{The Theory of Relativity}}}\ (\bibinfo {publisher}
  {Oxford University Press},\ \bibinfo {address} {Oxford},\ \bibinfo {year}
  {1952})%
  \bibAnnoteFile{NoStop}{moeller52}%
\bibitem{arzelies66}%
  \BibitemOpen
  \bibfield{author}{%
  \bibinfo {author} {\bibfnamefont{H.}~\bibnamefont{Arzeli\`es}},\ }%
  \emph{\bibinfo {title} {{Relativistic Kinematics}}}\ (\bibinfo {publisher}
  {Pergamon},\ \bibinfo {address} {New York},\ \bibinfo {year} {1966})%
  \bibAnnoteFile{NoStop}{arzelies66}%
\bibitem{cranor99}%
  \BibitemOpen
  \bibfield{author}{%
  \bibinfo {author} {\bibfnamefont{M.~B.}\ \bibnamefont{Cranor}}, \bibinfo
  {author} {\bibfnamefont{E.~M.}\ \bibnamefont{Heider}},\ and\ \bibinfo
  {author} {\bibfnamefont{R.~H.}\ \bibnamefont{Price}},\ }%
  \bibfield{title}{%
  \enquote{\bibinfo {title} {{A circular twin paradox}},}\ }%
  \bibfield{journal}{%
  \bibinfo {journal} {Am. J. Phys.}\ }%
  \textbf{\bibinfo {volume} {68}},\ \bibinfo {pages} {1016--1020} (\bibinfo
  {year} {1999})%
  \bibAnnoteFile{NoStop}{cranor99}%
\bibitem{Note14}%
  \BibitemOpen
  \bibinfo {note} {\protect \emph {Standard} clocks at \protect \emph
  {different} radii cannot be synchronized with each other this way.}%
  \bibAnnoteFile{Stop}{Note14}%
\bibitem{rizzi04a}%
  \BibitemOpen
  \bibfield{author}{%
  \bibinfo {author} {\bibfnamefont{G.}~\bibnamefont{Rizzi}}, \bibinfo {author}
  {\bibfnamefont{M.~L.}\ \bibnamefont{Ruggiero}},\ and\ \bibinfo {author}
  {\bibfnamefont{A.}~\bibnamefont{Serafini}},\ }%
  \bibfield{title}{%
  \enquote{\bibinfo {title} {{Synchronization gauges and the principles of
  special relativity}},}\ }%
  \bibfield{journal}{%
  \bibinfo {journal} {Found. Phys.}\ }%
  \textbf{\bibinfo {volume} {34}},\ \bibinfo {pages} {1835--1887} (\bibinfo
  {year} {2004})%
  \bibAnnoteFile{NoStop}{rizzi04a}%
\bibitem{Note15}%
  \BibitemOpen
  \bibinfo {note} {In practice this condition is satisfied, if the observers
  correspond to material points of a non-deforming solid.}%
  \bibAnnoteFile{Stop}{Note15}%
\bibitem{Note16}%
  \BibitemOpen
  \bibinfo {note} {If $\gamma $ is larger than $\protect \sqrt {2}$, more than
  two ants will simultaneously see a white mark.}%
  \bibAnnoteFile{Stop}{Note16}%
\end{thebibliography}

%Merlin.mbs v4.21 2009-07-09.
%

\end{document}